\documentstyle[epsfig]{aipproc}
\begin{document}
\def\thefootnote{\fnsymbol{footnote}}\advance\vsize by.5in
\rightline{ADP-00-49/M95; astro-ph/0010443}
\title{Supernovae Ia, Evolution and Quintessence\footnote{To appear in
{\it``Cosmology and Particle Physics''}, Proc.\ of the CAPP'2000
Conference, Verbier, eds.\ J. Garc\'{\i}a-Bellido, R. Durrer and M.
Shaposhnikov, (American Institute of Physics, 2001)}}
\author{David L. Wiltshire}
\address{Department of Physics and Mathematical Physics, University of
Adelaide,\\ Adelaide, S.A. 5005, Australia}
\maketitle

\begin{abstract}
Quintessence models with a dark energy generated by pseudo Nambu--Goldstone
bosons provide a natural framework in which to test the possibility that
type Ia supernovae luminosity distance measurements are at least partially due
to an evolution of the sources, since these models can have parameter values
for which the expansion of the Universe is decelerating as well as values for
which it is accelerating, while being spatially flat in all cases and
allowing for a low density of clumped matter. The results of a recent
investigation \cite{NW} of current observational bounds which allow
for SNe Ia source evolution are discussed. It is found that models with source
evolution still favour cosmologies with an appreciable amount of acceleration
in the recent past, but that the region of parameter space which is
most favoured shifts significantly.
\end{abstract}

%%%%%%%%%%%%%%%%%%%%%%%%%%%%%%%%%%%%%%%%%%%%%%%%%%%%%%%%%%%%%%%%%%%%%%%%%%%%
\def\goesas{\mathop{\sim}\limits} \def\LA{\Lambda} \def\etal{{\it et al.}}
\def\si{\sigma} \def\ph{\phi} \def\Om{\Omega} \def\w#1{\;\hbox{#1}\;}
\def\Omm{\Om_{m0}} \def\Omp{\Om_{\ph0}}
\def\Ommf{\Om_{m{\tt f}}} \def\Ompf{\Om_{\ph{\tt f}}}
\def\frac#1#2{{\textstyle{#1\over#2}}} \def\be{\beta} \def\br{\hfil\break}
%%%%%%%%%%%%%%%%%%%%%%%%%%%%%%%%%%%%%%%%%%%%%%%%%%%%%%%%%%%%%%%%%%%%%%%%%%%%

As has been described in many other talks at this conference, many
independent measurements -- for example, the spectrum of cosmic microwave
background radiation (CMBR) anisotropies, galaxy clustering statistics,
peculiar velocities, the baryon mass fraction in clusters of galaxies -- all
appear to agree in estimating that the density of matter which is
clumped is relatively low compared to the critical density, being of order
$\Omm\goesas0.2$--$0.3$.
On the other hand, the recent measurement of the position of the first
acoustic peak in the angular power spectrum of CMBR anisotropies by the
BOOMERANG-98 and MAXIMA-I experiments now gives unequivocal evidence
that the Universe is close to being spatially flat \cite{boomax}.

A natural conclusion to draw from these observations is that a significant
proportion of the energy density of the Universe is in the form of a dark
component which is smooth, rather than clumped, on cosmological scales.
The form of dark energy which we are most familiar with, for
historical reasons, is a cosmological constant. It leads to
models in which the expansion of the universe is accelerating.

In the last few years, Type Ia supernovae (SNe Ia) have come to be used as a
cosmological distance indicator, with the conclusion that there is
very good evidence that the expansion of the Universe is indeed
accelerating \cite{R98,P98}. Unfortunately, in terms of the physical basis of
the measurements, the SNe Ia result remains the most poorly understood
component of the present ``concordance model'', and it remains possible that
there are systematic uncertainties that have not been accounted for, such as
an evolution of the sources or extinction by dust. In particular, cosmological
parameters are fitted by normalizing the peak luminosities of supernovae on the
basis of a purely {\it empirical} correlation which has been observed at low
redshifts between the peak luminosity and the decay time of SNe Ia events as
measured in their rest frames: the resulting ``Phillips relations''
\cite{Ph93}--\cite{R96} reduce the dispersion in the distance moduli to about
$0.15$ \cite{Ham96,R96} as compared to an intrinsic dispersion of $0.3$--$0.5$
in peak absolute $B-V$ magnitudes of suitably selected nearby events.

\advance\vsize by-.5in \catcode`\@=11
\def\@evenfoot{\hss\tenrm--\ \number\thepage\ --\hss}
\let\@oddfoot\@evenfoot \catcode`\@=12
Many of the details of the physics behind SNe Ia remain unclear. It is believed
that each SNe Ia event is formed by a white dwarf in a binary
system, accreting matter from its companion until it undergoes a catastrophic
thermonuclear conflagration, possibly at a sub-Chandrasekhar limit stage.
Given their common physical origin such systems might be reasonably expected
to be somewhat ``insensitive'' to much of the individual histories of the
progenitor systems, which provides the basis for their use as a standard
candle. However, while much progress has been made in attempting to numerically
model the explosions \cite{blast}, huge uncertainties remain because of a lack
of knowledge of the details of particular nuclear cross--sections and the
fluid dynamics of flame propagation. Attempts to find a physical origin for
the Phillips relations are at a very preliminary stage \cite{PE00}.

Given that the prospect of understanding the physical basis of the SNe Ia
events is not going to be resolved without much more detailed measurements
and calculations, it would be prudent to test the conclusions that have been
derived cosmologically, allowing for the possibility that there has been
some evolution of the peak luminosities insofar as they affect the Phillips
relations over cosmological timescales. Such an analysis has been
instigated by Drell, Loredo and Wasserman \cite{DLW} in the case of open
Friedmann--Robertson--Walker (FRW) models.

I wish to argue, however, that since we now have remarkably good evidence
that the Universe is close to flat with a low density fraction of
ordinary clumped matter, $\Omm\goesas0.2$--$0.3$, it makes much better
sense to test the evolution hypothesis in the context of models which have
these features but which make no assumptions regarding the acceleration or
deceleration of the Universe at the present epoch. This is not possible
within the class of Friedmann-Lema\^{\i}tre models with $\Om_m+\Om_\LA=1$.
However, cosmological models with a quintessence field in the form of a
dynamical pseudo Nambu--Goldstone boson have precisely the desired properties.

It may come as a surprise that ``quintessence'' does not necessarily entail
an accelerated expansion of the Universe, since one of the most common
approaches to seeking a particle physics origin for the vacuum energy is to
look for a scalar field, $\ph$, with a potential, $V(\ph)$, which gives
homogeneous isotropic cosmological solutions for which the Universe undergoes
an accelerated expansion at late times. However, such an approach does not
fully utilize the fact that the effective equation of state for the
quintessence field, $P_\ph=w_\ph\rho_\ph$, has a {\it variable} coefficient
$w_\ph$ which can generically take all values consistent with the dominant
energy condition. Furthermore, such an approach often leads to the study of
potentials whose physical origin is not particularly well motivated.

The PNGB model, on the other hand, is well--motivated from a particle physics
point of view as being a very natural way of obtaining an ultra--light scalar
field \cite{pngb}. The models \cite{NW,pngb,FW} are based on a potential
$$V(\ph)=M^4[\cos(\ph/f)+1]$$
characterized by two mass scales, a
purely spontaneous symmetry breaking scale $f\goesas10^{18}$--$10^{19}\w{GeV}$,
and an explicit symmetry breaking scale $M\goesas10^{-3}\w{eV}$.

The solutions, for a model consisting of the quintessence field plus a
spatially flat homogeneous isotropic cosmology with clumped matter in the
form of dust, have the property that at late times the energy density fractions
in the scalar field and clumped matter tend to constants $\Ompf$ and $\Ommf$,
with $\Ompf+\Ommf=1$, while the scalar field oscillates about the minimum of
$V(\ph)$.
The values of $\Ompf$ and $\Ommf$ depend on the values of the parameters
$M$, $f$ and the initial value $\ph_i$ of the scalar field at the onset of
the matter domination. As the scalar field oscillates about the potential
minimum the deceleration parameter oscillates from a minimum value of
$q=\frac12\left(1-3\Ompf\right)$ to a maximum value $q=\frac12\left(1+3\Ompf
\right)$ about a mean of $\langle q\rangle=\frac12$.

After many oscillations it is the mean value $\langle q
\rangle$ which is significant, and the luminosity distance relation becomes
indistinguishable from that of the Einstein-de Sitter universe. Thus a
cosmological acceleration can have an appreciable effect at the present epoch
provided that (i) $\Ompf>\frac13$; and (ii) at the present epoch the scalar
field is still undergoing its first oscillation.

\begin{figure}[t!]
\centerline{\epsfig{file=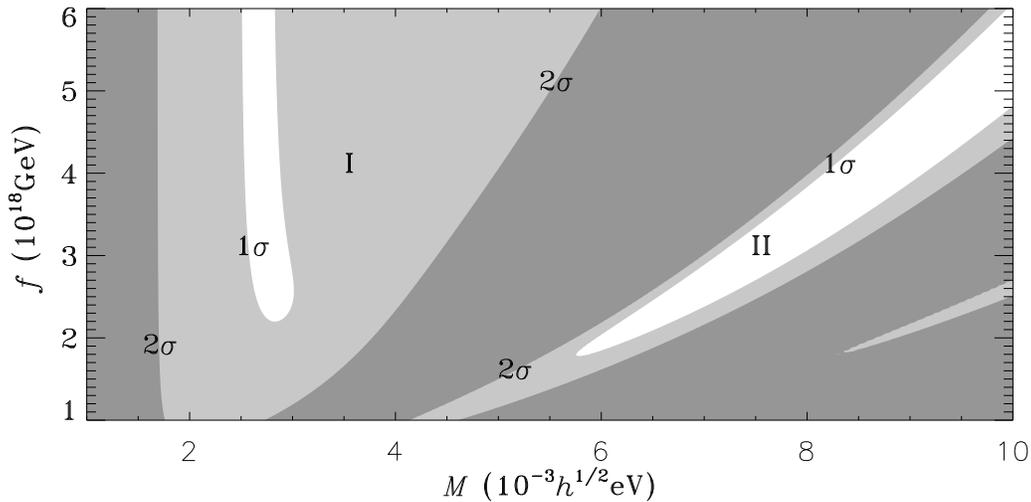,height=3in,width=6in}}
\vspace{10pt}
\caption{Confidence limits on $M,f$ parameter values, with $\ph_i=1.5f$, $\be_0
=0$, and $b=0.36$ (best--fit value), for the 60 SNe Ia in the Perlmutter
dataset [4]. Parameter values excluded at the 95.4\% level are darkly shaded,
while those excluded at the 68.3\% level are lightly shaded.}
\label{fig1}
\end{figure}

I now wish to briefly describe some of the results of recent work
\cite{NW}, in which Cindy Ng and I investigated the luminosity distance
confidence limits on the parameter space $(M,f,\ph_i)$ of PNGB quintessence
models, allowing for an additional continuous redshift--dependent magnitude
shift of the form $\be\ln(1+z)$ in the distance modulus relation for the SNe
Ia sample, as suggested by Drell, Loredo and Wasserman \cite{DLW}. The
parameter $\be$ is assumed to have a Gaussian prior distribution with mean
$\be_0$ and standard deviation $b$. We investigated two types of evolution:
(i) models with non-zero $\be_0$; (ii) models with $\be_0=0$ but non-zero
dispersion $b$ (see Fig.~\ref{fig1}). The former models
can be regarded as ones with a possibly strong evolution. In the second
case it is assumed that the Phillips relations still apply on average over
cosmological scales, but that evolutionary effects result in an additional
dispersion. This is potentially a weaker form of correction that may
apply to a variety of possible sources of evolutionary effects.

We determined confidence limits by analytic marginalization, using the
60 SNe Ia published by Perlmutter \etal\ \cite{P98}. We found that the models
with evolution provided a better fit to the data. In the case of type (i)
models with non-zero $\be_0$ the best fit value occurred at $\be_0=0.414$ for
$\ph_i=1.5f$, or $\be_0=0.435$ for $\ph_i=0.2f$. This corresponds to SNe Ia
being intrinsically dimmer by $0.17$--$0.18$ magnitudes at a redshift $z=0.5$.
The $\be_0=0$ slice of parameter space included regions which still lie within
the $2\si$ confidence region relative to the best-fit value of $\be_0$,
however, for all values of $b$. What is perhaps even more interesting is that
in the case of type (ii) models with
$\be_0=0$, if $\ph_i$ is suitably large -- e.g., for $\ph_i=1.5f$ as in Fig.\
\ref{fig1} -- a non-zero dispersion $b\simeq0.36$ can lead to a slightly better
fit than the case of non--evolutionary models with $b=0$.

\begin{figure}[t!]
\centerline{\epsfig{file=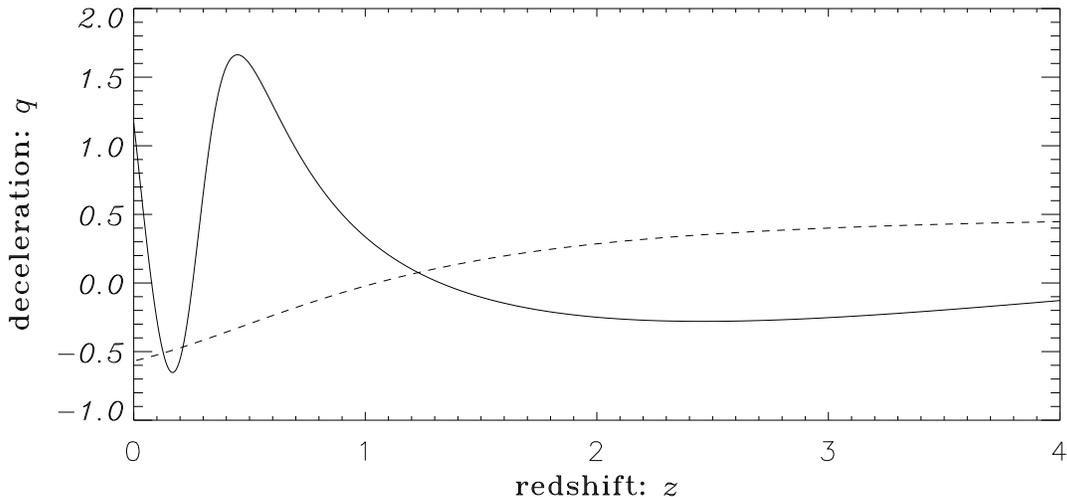,height=3in,width=6in}}
\vspace{10pt}
\caption{The deceleration parameter for two models with $\Omm\simeq0.2$ and
$\ph_i=1.5f$:\br (i) the dashed line shows $q(z)$ for $M=0.0029\;h^{1/2}\w{eV}
$, $f=3.5\times10^{18}\w{GeV}$, a point in region I;\br (ii) the solid line
shows $q(z)$ for $M=0.0066\;h^{1/2} \w{eV}$, $f=2.3\times10^{18}\w{GeV}$, a
point in region II.}
\label{fig2}
\end{figure}

The assignment of $1\si$ and $2\si$ confidence bounds of course depends on
the dispersion, $b$, and thus the conclusions one can draw at this stage will
remain somewhat qualitative until a better physical understanding of any
evolutionary effects is obtained. What is perhaps most important is that for
type (ii) models with $b>0.17$ most of the $1\si$--included region is located
in the region marked II in Fig.\ \ref{fig1}, in which the scalar field has
already passed through its minimum once and is rolling back towards the
minimum for the second time. By contrast, for the non-evolutionary case, $b=0$,
the $1\si$--included region is {\it entirely} within in the area marked I (see
Fig.\ 8 of \cite{NW}), in which
the scalar field is rolling down the potential for the first time at the
present epoch without yet having reached the minimum. Region I corresponds
to parameter values for which the conventional quintessence scenario applies,
since the deceleration parameter is negative (c.f.\ Fig.\ \ref{fig2}). For
the strongly evolving type (i) models the picture is similar, except that
relative to the best--fit $\be_0$ {\it all} of region I is $1\si$--excluded
for values $b\le0.5$ (c.f.\ Figs.\ 9--11 of\cite{NW}).
For parameter values in region II it is much easier to
accommodate gravitational lensing statistic constraints from
quasars than for parameter values in region I \cite{NW}.

In region II, although the deceleration parameter is positive at the present
epoch, the universe would have experienced a significant amount of acceleration
at modest redshifts within the SNe IA dataset, as can be seen from the plot of
$q(z)$ in Fig.~\ref{fig2} for some typical parameter values.
It is interesting therefore that even with the inclusion of source
evolution the SNe Ia data best fits parameter values for which the cosmological
evolution differs markedly from that of open FRW models. With future data from
the proposed SNAP mission \cite{PNR} it would become possible to
observationally distinguish region II PNGB quintessence models from
Friedmann-Lema\^{\i}tre models and other quintessence scenarios.

\smallskip
{\large\it Acknowledgement:}\quad I wish to thank Cindy Ng for discussions.
This work was supported by Australian Research Council grant F6960043.
\def\PRL#1{{\it Phys.\ Rev.\ Lett.\ {\bf#1}}}
\def\PR#1{{\it Phys.\ Rev.\ {\bf#1}}}
\def\ApJ#1{{\it ApJ\ {\bf#1}}}
\def\AsJ#1{{\it Astron.\ J.\ {\bf#1}}}

\end{document}